\documentclass[sigconf]{acmart} 

\AtBeginDocument{%
  }

\begin{document}

\title{GenPod: Constructive News Framing in AI-Generated Podcasts More Effectively Reduces Negative Emotions Than Non-Constructive Framing}

\author{Wen Ku}
\authornote{Both authors contributed equally to this research.}
\affiliation{%
  \institution{Tsinghua University}
  \city{Shenzhen}
  \state{Guangdong}
  \country{China}
}

\author{Yihan Liu}
\affiliation{%
  \institution{Renmin Univesity of China}
  \city{Beijing}
  \country{China}}

\author{Wei Zhang}
\affiliation{%
  \institution{Shenzhen University}
  \city{Shenzhen}
  \state{Guangdong}
  \country{China}
}

\author{Pengcheng An}
\affiliation{%
 \institution{Southern University of Science and Technology}
 \city{Shenzhen}
 \state{Guangdong}
 \country{China}}


\begin{abstract}
AI-generated media products are increasingly prevalent in the news industry, yet their impacts on audience perception remain underexplored. Traditional media often employs negative framing to capture attention and capitalize on news consumption, and without oversight, AI-generated news could reinforce this trend. This study examines how different framing styles—constructive versus non-constructive—affect audience responses in AI-generated podcasts. We developed a pipeline using generative AI and text-to-speech (TTS) technology to create both constructive and non-constructive news podcasts from the same set of news resources. Through empirical research (N=65), we found that constructive podcasts significantly reduced audience's negative emotions compared to non-constructive podcasts. Additionally, in certain news contexts, constructive framing might further enhance audience self-efficacy. Our findings show that simply altering the framing of AI generated content can significantly impact audience responses, and we offer insights on leveraging these effects for positive outcomes while minimizing ethical risks.
\end{abstract}

\begin{teaserfigure}
  \includegraphics[width=\textwidth]{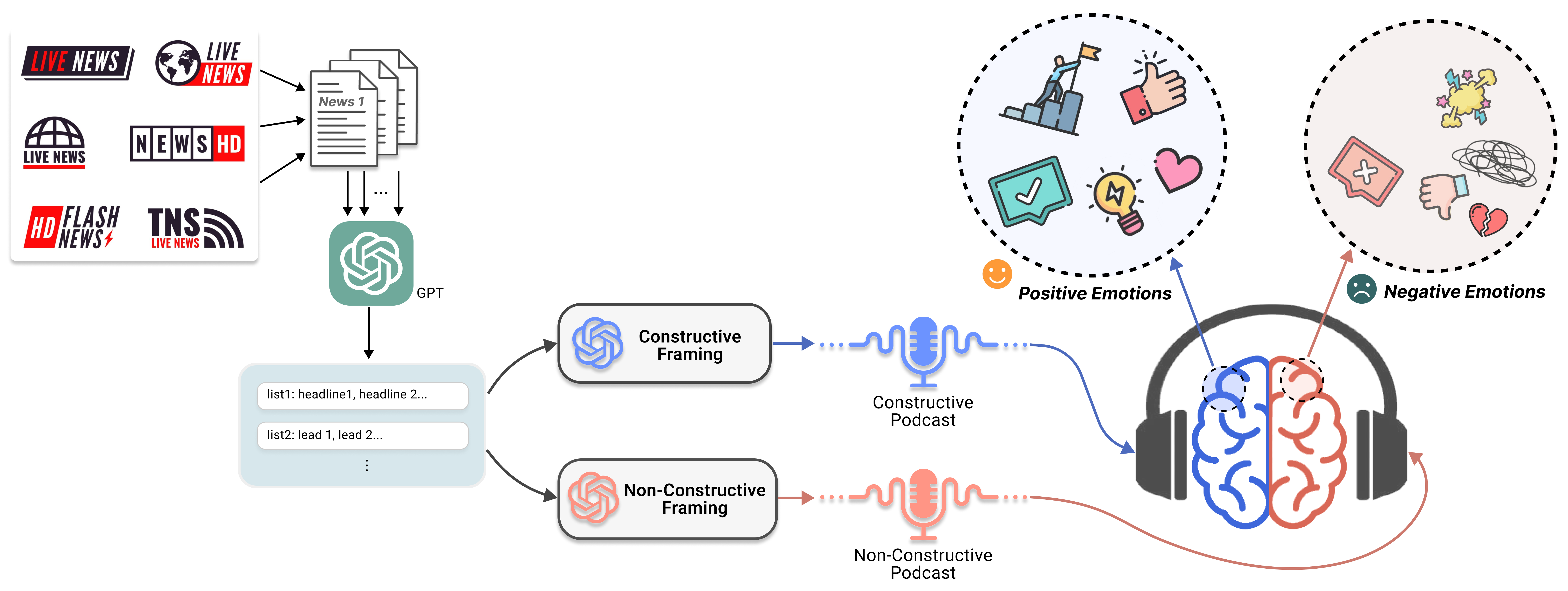}
  \caption{An illustration of the generative podcast production pipeline with constructive or non-constructive framing methods and their different influences on emotions. A more comprehensive process description is given in Section 4.}
  \Description{Enjoying the baseball game from the third-base
  seats. Ichiro Suzuki preparing to bat.}
  \label{fig:teaser}
\end{teaserfigure}


\maketitle

\section{Introduction}
With the rapid advancement of artificial intelligence, AI-driven media products are becoming an increasingly integral part of everyday life. Recently, limited research has started exploring the emerging medium of AI-generated podcasts. However, the effects of such content remain insufficiently understood, particularly regarding the safe use of these technologies to avoid associated risks. Research in traditional media has demonstrated that the framing of news narratives significantly influences public perception and behavior. 

Negative framing is a common approach in news media, favored for its ability to drive audience engagement. Research shows that negative news tends to be more attention-grabbing and easily shared, which contributes to its widespread reach \cite{Nabi2003Exploring}. In light of this, our study compares two framing approaches: constructive and non-constructive (negative) news framing. Constructive framing emphasizes solutions and forward-looking perspectives, offering a potential way to mitigate the negative psychological impacts linked to traditional news reporting.

This study addresses the question: \textbf{How do audience responses differ when the same news resources are compiled into podcasts by an LLM using either constructive or non-constructive framing?} To investigate this, we developed a generative AI pipeline integrated with text-to-speech (TTS) technology. The pipeline takes identical news materials as input and, through prompt engineering, produces both constructive and non-constructive podcast formats with the same materials. We then conducted an empirical study to compare the effects of these AI-generated podcasts on listeners' emotions and self-efficacy.

The empirical results indicate that, compared to non-constructive news, constructive podcasts can somewhat alleviate listeners' negative emotions and enhance their self-efficacy. The study makes two main contributions. First, it provides empirical data to reveal how AI-generated constructive and non-constructive news podcasts influence listeners' emotions and self-efficacy. Second, it offers a set of implications on how AI-generated podcasts can be positively leveraged in HCI system design, while also addressing ways to mitigate potential ethical risks.

\section{RELATED WORK}

\subsection{AI-Generated News and Its Challenges}

In recent years, AI-generated news products have rapidly gained prominence in the media industry, becoming a key component of news production. Numerous studies have explored AI's potential in automating news production \cite{visvam2019efficient,lopezosa2023use,ioscote2024artificial} and personalizing content \cite{fu2023ai,zhang2024heterogeneous}, highlighting its role in improving efficiency and accuracy \cite{torkamaan2024role}. However, as AI technology becomes more integrated into journalism, research has also revealed potential risks such as generating deceptive content \cite{sandrini2023generative}. This can be fake news \cite{gupta2022combating}, tabloid-like clickbait \cite{xu2019clickbait}, or biased content on social conflict issues like race, religion, gender, and class \cite{nazar2020artificial,woolley2020we}. Xiao’s research indicates that even reputable outlets like The New York Times and Reuters may produce AI-generated content with notable gender and racial biases \cite{fang2024bias}.

These findings underscore the risks associated with AI in news generation, extending beyond content bias or misinformation. Ensuring the transparency and reliability of AI systems has become a pressing issue, especially given the "black box" nature of  AI systems \cite{adadi2018peeking,mitova2023exploring}, which can lead to fact misrepresentation \cite{mahmood2023fact} or the creation of misleading content, known as "hallucinations" \cite{almeida2022podcast,bruno2023insights}. Researchers are actively exploring methods to address these challenges, focusing on enhancing AI transparency \cite{zhou2020towards,mitova2023exploring} and reliability \cite{kim2024m,du2023personalization}. Discussions in HCI increasingly consider the ethical risks of generative AI in news production \cite{xu2023transitioning,shi2023hci,trattner2022responsible}, emphasizing both technological improvements and responsible usage \cite{aissani2023artificial,gutierrez2023ai}.

Although significant efforts have been made to ensure the accuracy and reliability of AI-generated content, another critical aspect remains underexplored in HCI: the choice of framework for presenting facts, while keeping the underlying news material and facts unchanged \cite{caramancion2023news,dierickx2023ai}.

\subsection{Impact of News Framing on Consumer Cognition and Behavior}

Outside of HCI community, in the field of journalism, news framing significantly influences consumer responses. Research shows that negative news frames often provoke strong emotional reactions, such as fear or anger, which can heavily impact public opinion and behavior \cite{Entman1993Framing,Scheufele1999Framing,Nabi2003Exploring}.  This tendency is further amplified by commercial pressures within the traditional news industry, which often favors negative framing due to its higher potential for virality \cite{pedersen2014news} and attention \cite{trussler2014consumer,van2019mediatization}, as well as its ability to provoke anxiety \cite{ricciardelli2024news}, concern \cite{pedersen2014news,soley1992advertising} and even cynicism \cite{Baden2019Impact} among audiences.

In contrast, positive or constructive frames, focusing on solutions and future perspectives, promote the audience's optimism and self-efficacy \cite{Gyldensted2015Mirrors,Van2022Effects,Overgaard2023Mitigating}. Constructive journalism, which emphasizes solution-oriented narratives, has been shown to reduce the negative psychological impacts of traditional news, foster balanced emotional responses, and encourage proactive behaviors \cite{Baden2019Impact,Gyldensted2015Mirrors}. Research suggests that constructive news not only reduces anxiety but also enhances self-efficacy and motivation to engage with social issues \cite{Diakopoulos2019Automating,McIntyre2018Reconstructing}, thereby fostering prosocial behavior \cite{Van2022Effects,Kleemans2017Preadolescents}.

Despite the growing research on the impact of news framing and constructive journalism, the role of AI in shaping news frames and its subsequent effects on audience perception remain underexplored \cite{Carlson2018Robotic,Bucher2018If}. There is a concern that AI-driven news production could replicate the commercial biases of traditional journalism, potentially prioritizing negative framing to capture attention \cite{thurlow2020visualizing,lee2022neus}. As AI's role in news production grows, it is crucial to examine how AI-generated news, whether framed constructively or non-constructively, affects audiences. Our study seeks to address the critical gap.

\subsection{AI-Generated Podcast}

Podcasts, as an emerging digital media form, offer a rich diversity of language and expression, catering to the fast-paced lifestyle of modern audiences who often consume information aurally while multitasking. Podcast is particularly popular among younger audiences, indicating its potential as a future trend in news consumption. Consequently, we chose podcasts as the starting point to explore how AI-generated news, framed either constructively or non-constructively, influences listener responses.

Although AI's application in news generation is becoming increasingly widespread, research in the podcast domain primarily focuses on optimizing recommendation algorithms \cite{aluri2023optimizing,liang2023enabling,nazari2022choice,zhang2024simulating} and extracting summaries \cite{karlbom2021abstractive,vaiani2022leveraging,rezapour2022makes}. However, recent years have seen the emergence of various applications in industry and practice. Prior to the widespread use of large language models (LLMs), some studies explored the use of natural language processing (NLP) to generate podcast content \cite{almeida2022podcast} or streamline the podcast creation process \cite{rime2022you}. Recent industry developments, such as Podcastle \cite{podcastle2024} and Beyondwords \cite{beyondwords2024}, have begun incorporating Generative AI (GenAI) and text-to-speech (TTS) technologies into podcast production. Despite their advancements, these tools are still in the early stages and lack a robust research framework that supports news automation.

To address these gaps, we developed a pipeline using LLMs to structure news content with constructive and non-constructive framings, allowing us to compare their effects on listener responses.

\section{FORMATIVE STUDY}

We conducted a formative study with two goals. Firstly, to gain a vivid understanding of podcast listeners' current listening practices. Secondly, to obtain podcast users' specific expectations for generative podcasts, and to understand the essential user needs to inform our design procedure.

\subsection{Participants and Procedure}

We recruited 8 regular podcast listeners (F1-8, 4 , 4 male; age, M = 24.25, SD = 3.03) through snowball sampling, with listening experience ranging from 3 months to over 2 years.

Two 2-hour online workshops were held, each with 4 participants. We created three generative podcast samples using Large Language Models (LLM) and Text-To-Speech (TTS) tools as design probes to gather feedback. Each sample lasted 1.5 to 2 minutes and was created referring to existing podcast themes and styles.

Workshops began with a 30-minute discussion on listening habits, followed by a video explaining the generative podcast production. Participants then listened to the design probes and engaged in a 50-minute discussion on their listening experiences. Finally, a 40-minute co-design activity had participants create their own generative podcast and explain their design choices.

\subsection{Formative Study Insights}

Participants demonstrated a versatile interest in podcasts. They listened in relaxed as well as busy contexts such as during commutes, household chores, or while coding, illustrating the adaptability and ubiquitous nature of podcasts. The key insights we have learned from the participants are twofold:

\subsubsection{News podcasts were deemed a suitable format to be generated by current AI}

Participants believed AI could efficiently gather diverse news material. \emph{"It's convenient for accessing various real-time events,"}(F2) \emph{"The time for AI to gather fresh news is much lower than humans."}(F5) AI was also perceived as capable of delivering more objective perspectives. \emph{"AI has no personal bias, so it won’t selectively overlook or favor certain viewpoints"}(F6), which \emph{"makes it sounds professional and credible."}(F3) However, participants also recognized AI's limitations in conveying emotion, \emph{"The tone is rather flat and straightforward,"}(F1) \emph{"makes it more suitable for reporting emotion-neutral news."}(F4) Overall, participants agreed that news was an ideal format for generative podcasts, aligning with the focus of our research——to develop a pipeline for AI to compile and present news content.

\subsubsection{Dyadic dialogues were considered more engaging for generative news podcast}

Participants found conversational formats help to create a sense of connection. \emph{"AI can connect more emotionally with a dialogue, as it mimics real-life conversations,"}(F2) The dynamic nature of two-person interactions was also highlighted. \emph{"You can feel the interaction between the hosts, which brings out dynamic."}(F5) Additionally, dialogue was seen as well-suited for presenting differing perspectives. \emph{"The other person can repeat or ask questions, which helps clarify and emphasize a point."}(F6) and \emph{"makes it easier to process because the back-and-forth breaks down dense information."}(F7) Moreover, the dialogue could \emph{"reduce the artificiality of AI voices and alleviate monotony"(F3) as "it shifts attention away from the voice itself"}(F4). For these reasons, dyadic dialogue was considered an effective way to enhance the overall podcast experience.

\begin{figure*}[h]
  \centering
  \includegraphics[width=\linewidth]{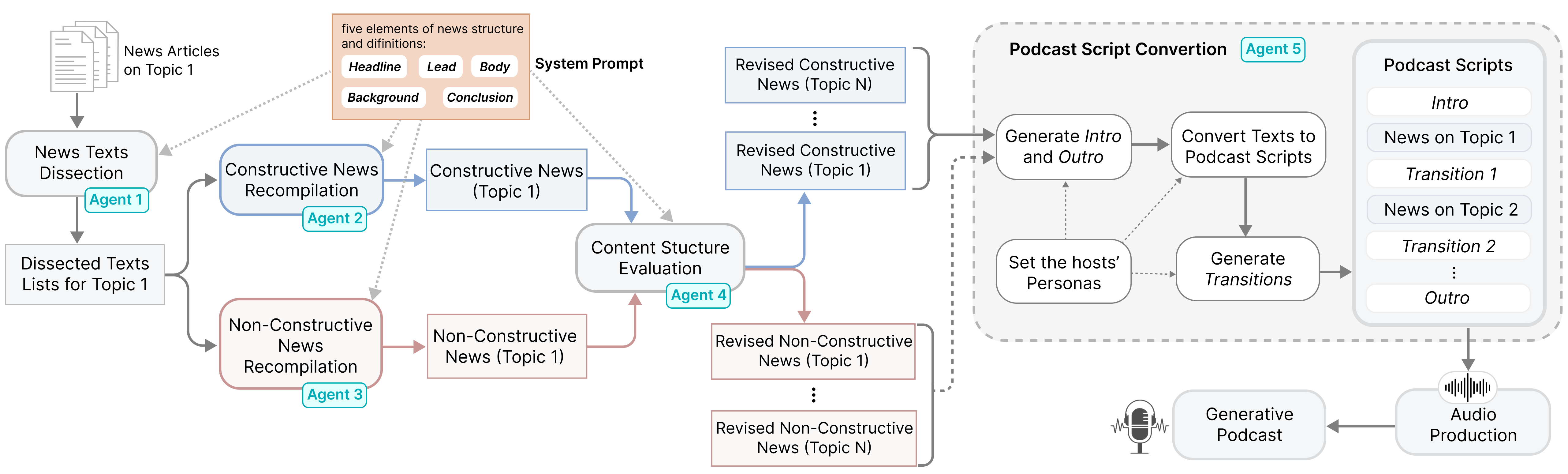}
  \caption{Overview of GenPod’s pipeline to generate constructive and non-constructive podcasts. The initial input consists of news on the same topic, sourced from authoritative news platforms, with the final output being two types of podcasts: constructive and non-constructive. The process involves five agents in sequence, while the System Prompt serves as a foundation throughout the former four agents. Agent 1 dissects news sources into basic components: Agents 2 and 3 recompile these components using constructive or non-constructive frames. Agent 4: evaluate whether the recompiled content Agent 4 evaluates whether the recompiled content meets the requirements of news structure. Agent 5 transforms the plain text into podcast scripts. Finally, (Text-To-Sound)TTS technology and post-production sound effects are used to produce the audio.}
  \Description{Overall Pipeline}
  \label{fig:2}
\end{figure*}

\section{GenPod: A Generative AI Pipeline for News Podcasts}

We aim to probe how audience responses differ when AI-compiled news podcasts use either constructive or non-constructive framing. To do so, we developed GenPod, an AI-based pipeline that creates news podcasts by drawing from multiple reputable news platforms to produce two podcasts, one framed constructively and one non-constructively.

The pipeline involves four stages, as shown in Fig. \ref{fig:2}:
(1) \textit{Dissecting news sources}: where the news sources were dissected into basic components.
(2) \textit{Recompiling News}: where the components were recompiled into plain texts framed differently. 
(3) \textit{Podcast script conversion}: where the plain texts were transformed into podcast scripts.
(4) \textit{Audio production}: where the scripts were converted into audio products.

\subsection{Dissecting news sources from reputable news platforms}

GenPod uses a system prompt based on a standard news structure, which includes five elements: headline, lead, body, background, and conclusion \cite{van1985structures,robescu2008investigating} and their definitions, serving as a foundation throughout the process.

\begin{figure*}[h]
  \centering
  \includegraphics[width=\linewidth]{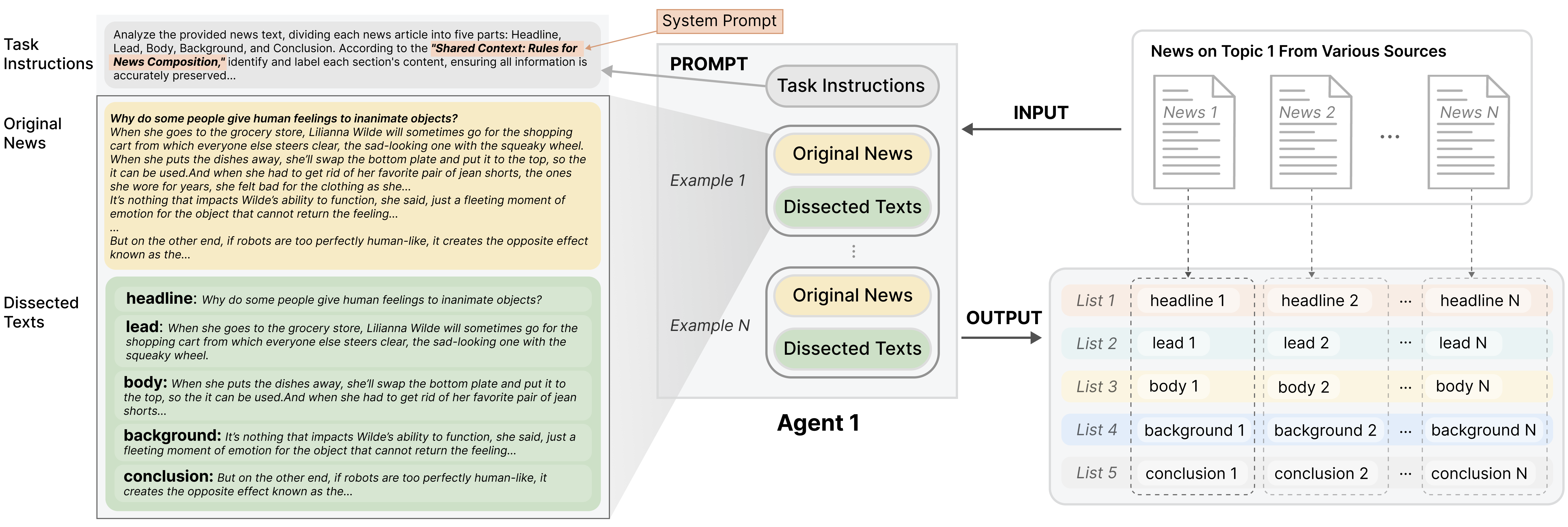}
  \caption{The detailed prompt structure and the function of Agent 1. The prompt for Agent 1 includes Task Instructions and several examples.  explicitly guide the agent to dissect each original article into five components: headline, lead, body, background, and conclusion, without any alterations. This is based on a system prompt that defines and explains each of these five elements. Examples demonstrate some news articles alongside their corresponding dissection results. When multiple news articles on the same topic are input, Agent 1 dissects them and categorizes the results by element type, ultimately outputting five lists, serving as the input for the subsequent phase.}
  \Description{The detailed prompt structure and the function of Agent 1}
  \label{fig:3}
\end{figure*}

Agent 1 analyzes articles on the same topic from various reputable sources, labeling and dissecting them according to these elements. Notably, Agent 1 is instructed not to alter, summarize, or edit the original text, solely dissect the texts. Few-shot examples are provided to facilitate its understanding and execution of the task. The output is five consolidated lists of news elements, serving as the input for the subsequent phase. The detailed prompt structure and output of Agent 1 are shown in Fig. \ref{fig:3}.

\subsection{Recompiling News with Constructive and Non-Constructive Frames}

\subsubsection{Theoretical Basis}

To enhance the Large Language Model's (LLM) understanding of constructive and non-constructive news, we clearly outlined their definitions and key characteristics. Constructive news incorporates positive psychology, aiming to produce engaging content while fulfilling the core functions of journalism \cite{mcintyre2018constructive}. It focuses on potential solutions and promotes positive emotions \cite{mast2019constructive}. In contrast, non-constructive news emphasizes negative events and conflicts, deliberately excluding solutions or positive outlooks \cite{Van2022Effects}, often adopting a pessimistic tone without offering solutions \cite{Baden2019Impact}.

\subsubsection{Technical Implementation}

Based on definitions, GenPod utilized Agent 2 and 3, each incorporating characteristics of constructive and non-constructive news. The agents recompile the news content using the structured lists generated earlier, retaining the original details to avoid excessive compression that could compromise the quality. Representative few-shot examples \cite{worldsbestnews_womenrights,cnn_ai_jobs} were included for both types of news, which exhibit key characteristics of respective types. To ensure that the output adhered to the news structure criteria, Agent 4 evaluated and suggested modifications based on the system prompt.

All prompts and outputs were reviewed by two journalism experts. They confirmed the accuracy of the prompts and validated the quality and relevance of the generated news.

\subsection{Podcast Script Convertion}

Using news texts generated above, GenPod employed Agent 5, based on the Earkind framework \cite{earkind2024}, to convert them into podcast scripts. Earkind is an AI-driven system that generates podcast scripts based on user inputs. GenPod customized its prompt design to create tailored news podcast scripts. The system adopted a dyadic-dialogue narrative style, assigning specific roles to two hosts to enhance comprehension \cite{ginns2010personalization}, then converted the plain texts into podcast scripts, tailored to the characteristics of the two hosts. Finally, introductions, conclusions, and transitions between news segments were generated, ensuring a natural dialogue.

\subsection{Audio Production}

We used GPT-SoVITS, a speech synthesis model optimized for voice cloning, to generate the audio. The technology can closely mimic human speech and simulate natural conversational flow \cite{ijiga2024harmonizing,zhang2023does}. We developed both male and female voice models, alternating them in reading the scripts to simulate a conversation. Finally, we integrated the voice segments, adding appropriate pauses, silences, and background music with fade-ins and fade-outs, to create a complete podcast.

\section{METHODOLOGY}

The study investigates the impact of podcasts generated using the pipeline described in Chapter 4 on listeners' emotions and self-efficacy, from both constructive and non-constructive perspectives. To explore this question, we conducted a mixed-methods study, incorporating both quantitative and qualitative research approaches.

\subsection{Participants}

We recruited 66 participants (referred to as P1-66) through social media advertisements and snowball sampling (34 self-identified males, 31 self-identified females, 1 self-identified non-binary; aged 18-51, M = 23.23, SD = 5.420). Participants were invited to take part in an online study.

\subsection{Preparation of Study Materials}

Two topics were selected for the experiment: the challenges faced by food delivery workers and the controversies surrounding sports fandom during the Olympics. These topics were chosen due to their current relevance and inherent negative aspects, which made them suitable for adaptation into constructive perspectives. To ensure the AI had sufficient material for well-rounded content generation, we sourced several highly relevant articles on each topic from reputable news websites. The selection process aimed to balance perspectives and reduce potential bias by including both constructive elements (e.g., solutions and positive outcomes) and non-constructive elements (e.g., problems and negative outcomes).

Using the podcast generation pipeline in Chapter 4, we input these articles related to two specific topics to produce a constructive angle podcast episode (3:46 minutes) with each topic segment lasting between 80 to 90 seconds. The same method was applied to produce one episode with a non-constructive angle (3:44 minutes).  Consistency was maintained across both versions by the same introductory and concluding text, identical voices, and background music. The podcasts were uploaded to one of the largest podcast platforms. To enhance participant focus, the show notes included a detailed transcript and a timestamped outline.

\subsection{Procedure}

Participants first completed a pre-experiment questionnaire to assess their current emotional state using a 10-item Positive and Negative Affect Schedule (PANAS) scale \cite{watson2007development}. They were then randomly assigned to one of two groups: one group listened to the Constructive Podcast(CP) (P1-33), while the other listened to the Non-Constructive Podcast(NP) (P34-66). Participants were instructed to concentrate on the news content, minimizing potential distractions. After listening, they completed a post-experiment questionnaire, which included a comprehension check to ensure they had fully engaged with the podcast (e.g., summarizing the theme of the second news piece). The PANAS scale was re-administered to measure post-experiment emotional state. Three custom questions, tailored to each opic, were used to assess self-efficacy \cite{bandura1977self}. All questions were measured using a 7-point Likert scale.

Following the survey, semi-structured interviews were conducted with a randomly selected subset of seven participants to gather detailed insights to complement the quantitative data and explore the reasons for participants' responses. During the interviews, participants listened to the podcast version they had not heard during the experiment. They compared the two, discussing how each affected their emotions and self-efficacy and providing possible explanations. Participants were encouraged to share any additional insights or feedback. All interviews were audio-recorded, transcribed verbatim, and analyzed thematically. The questionnaire for this study received ethical approval. Participants were compensated at a rate of approximately 10 dollars per hour.

\subsection{Data analysis}

We categorized the levels of feeling alert, inspired, determined, attentive, and active from the PANAS scale as positive emotions, while the levels of feeling upset, hostile, ashamed, nervous, and afraid were classified as negative emotions. 

We analyzed the interview data using a thematic approach to develop codes and themes \cite{braun2006using}. The initial coding scheme was developed through an inductive process, emerging directly from the data. Two researchers collaborated to annotate the data and create the coding scheme. After the initial annotation and scheme development, the two researchers began independently coding the transcription data. Upon completing the independent coding, an iterative review process was conducted, consolidating and reconciling the coded datasets. This process included in-depth discussions to address discrepancies and collaboratively refine the coding results.

\section{FINDINGS}

\begin{figure}[h]
  \centering
  \includegraphics[width=\linewidth]{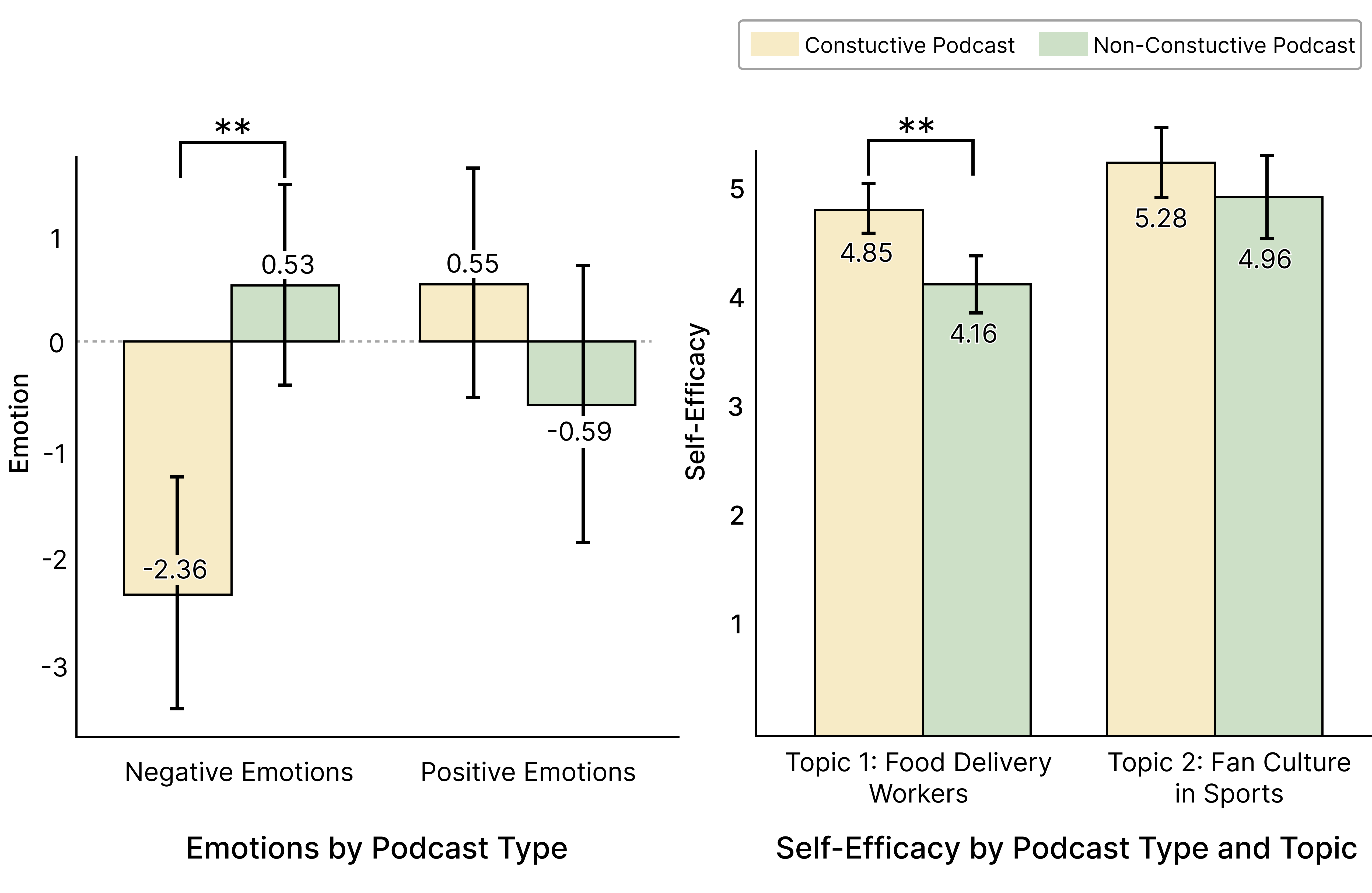}
  \caption{Results of the post-survey showed that constructive podcasts significantly reduced negative emotions and enhanced self-efficacy on topic 1: food delivery workers. ** indicate significance of $p$ < 0.01.}
  \Description{qualititive results}
  \label{fig:4}
\end{figure}

Among the 66 questionnaires collected, one was excluded due to incorrect responses in the integrity check, leaving 65 valid questionnaires. The change in emotional state was calculated as the difference between post-experiment and pre-experiment measurements.

\subsection{The Effect of Framing on Emotions}

\subsubsection{Negative Emotions}

ANOVA results (Fig. \ref{fig:4}) indicated a significant difference in the impact of Constructive Podcasts (CP) and Non-Constructive Podcasts (NP) on negative emotions ($F(1,63)=7.815, p<0.01$). Specifically, CP ($M = -2.36, SD = 4.46$) reduced negative emotions more than NP ($M = 0.53, SD = 3.85$).

\subsubsection{Positive Emotions}

There was no significant difference in positive emotions between CP and NP ($CP: M = 0.55, SD = 4.44; NP: M = -0.59, SD = 5.31$).

\subsubsection{Qualitative Results}

Qualitative interviews highlighted the contrasting emotional impacts of Constructive Podcasts (CP) and Non-Constructive Podcasts (NP) on participants. The distinct influence of these news types on emotions emerged from the discussions.

NP was linked to feelings of anxiety, depression, and helplessness due to its focus on severe and complex societal issues. After comparing the two types of podcasts, P3 noted, "\emph{about the delivery worker struggling to make ends meet, NP presented the situation as a dead-end, leaving no clear path forward}," while P5 added, "\emph{NP's detailed depiction of sports fan culture depicted it as a toxic environment with no hope for change.}" NP was criticized for its critical tone and lack of objectivity. P4 shared, "\emph{CP's balanced presentation, especially on delivery workers, made it more objective and comfortable to listen to.}"

In contrast, CP was appreciated for addressing negative societal issues while providing solutions and promoting positive actions, which helped alleviate negative emotions. P1 mentioned, "\emph{CP discussed both workers' struggles and new policies being implemented, showing that actions are being taken}." P6 added, "\emph{It wasn't just about what's wrong, but also about what's being done to make things better.}"

CP was noted for its ability to evoke calmness, objectivity, and hope. P3 remarked, "\emph{the success stories of other delivery workers who managed to improve their situations made me feel like there was a path forward, something to aspire to.}," P5 reflected, "\emph{CP's sports fandom report highlighted efforts to reduce toxicity, giving me hope that the situation could improve}." On the other hand, NP's focus on problems led to less positive emotional responses. P1 explained, "\emph{It only discusses the severity of their struggle, making it hard to feel motivated}," P6 pointed out, "\emph{Even when discussing solutions, NP emphasizes their negative consequences}."

\subsection{Self-Efficacy}

\subsubsection{Topic 1: Food Delivery Workers}

As to the topic of food delivery workers, self-efficacy differed significantly between conditions ($F(1,63) = 8.530, p < .01$), with CP ($M = 4.85, SD = 0.90$) higher than NP ($M = 4.16, SD = 1.01$).  (Fig. \ref{fig:4})

\subsubsection{Topic 2: Fan Culture in Sports}

For the topic of fan culture in sports, there was no significant difference between CP ($M = 5.28, SD = 1.27$) and NP ($M = 4.96, SD = 1.51$).

\subsubsection{Qualitative Results}

Participants who listened to the Constructive Podcasts (CP) about the challenges faced by food delivery workers generally reported feeling more empowered and capable of contributing to positive change. P2 noted, "\emph{Hearing about new regulations like setting minimum wages and ensuring safety standards made me believe that local advocacy could really make a difference.}" P3 added, "\emph{Knowing that there are already actionable steps in place made me feel like this issue is not just addressable, but that progress is already happening.}" The depictions of solutions and steps enhanced participants' confidence in their ability to effect change.

Conversely, listeners of Non-Constructive Podcasts (NP) expressed feelings of passivity. P2 remarked, "\emph{NP just highlights the difficulties faced by food delivery workers, like lack of job security and the high risk of accidents, without offering hope, leaving me feeling powerless.}" P5 similarly noted, "\emph{as if there was nothing that could be done to make things better.}" This sense of helplessness likely contributed to lower self-efficacy.

In the context of sports fan culture, qualitative data revealed that regardless of podcast type, participants felt a lack of control over the issue. P3 perceived the problem as "\emph{deeply entrenched and beyond the reach of individual efforts.}" P7 stated, "\emph{Even though CP tried to present strategies like educational campaigns, it doesn't change the fact that fan behaviors are something we cannot significantly influence.}"

Even those who listened to CP did not report increased self-efficacy regarding sports fan culture. The primary reason was that the proposed solutions in CP were seen as too idealistic. P6 observed, "\emph{While promoting positive role models were suggested for guiding fans, these ideas seemed unrealistic in practice.}"Moreover, P5 felt that "\emph{the positive tone of CP, while well-intentioned, made the issue seem less serious. It felt like they were downplaying the severity of the problem."} 

\section{DISCUSSION}

In the context of news and emotion research, prior studies have demonstrated that constructive media content can effectively regulate audience emotions \cite{Van2022Effects,Overgaard2023Mitigating}. However, most research has focused on visual media like video and text, with little attention to audio formats such as podcasts, which are increasingly popular. To address this gap, we employed prompt engineering to modify the presentation of existing news materials using LLM models, generating two podcast types: constructive and non-constructive. We then examined their effects on user emotions and self-efficacy.

We found that non-constructive podcasts(NP) significantly increased users' negative emotions, while constructive podcasts(CP) helped reduce them, supporting prior visual media research\cite{mcintyre2019solutions} and highlighting audio media's unique potential in emotion regulation. Although quantitative data showed no significant differences in positive emotions between podcast types—a common media effects research outcome \cite{wilson1993source}—qualitative data suggest CP may enhance positive emotions. According to the broaden-and-build theory \cite{fredrickson2001role}, positive emotions and solutions in CP could improve listeners' emotional states. Prior research indicates these positive changes become more pronounced with repeated or prolonged exposure \cite{mares2012effects}.

Regarding self-efficacy, while constructive podcasts enhanced users' self-efficacy in some cases, the overall effect was less significant than expected, aligning with mixed results from previous studies \cite{curry2014power,dahmen2021creating,mcintyre2015constructive}. Differences in self-efficacy may arise from: 1) the relatable nature of the delivery workers' stories versus the more abstract Theme 2; 2) variations in news content complexity and relevance; and 3) listeners' differing topic interests.

\subsection{Design Implications}

\subsubsection{AI-generated media has significant potential to positively impact society by employing constructive framing.}

Our research demonstrates the significant potential of generative AI in influencing audience emotions and self-efficacy by merely adjusting the constructiveness of news content. This aligns with existing studies on media framing effects\cite{mishra2023role} and reveals design opportunities for utilizing AI-generated media to benefit society.

For example, AI-generated media could be designed to foster positive emotions like hope and resilience, benefiting therapeutic contexts where individuals face stress or anxiety \cite{gual2022using}. Additionally, AI-generated content could raise awareness and motivate action on crucial social issues such as environmental sustainability \cite{goralski2020artificial} and equality \cite{mishra2023role}, by emphasizing constructive narratives and practical solutions. This approach can help audiences recognize the importance of the topics and empower them to contribute to meaningful change, echoing Entman’s findings on media's power in shaping public behavior \cite{Entman1993Framing}.

Beyond individual well-being, AI-generated content holds promise for broader social impacts. AI could dynamically tailor content to users' emotional states, offering supportive narratives during challenging times. Prioritizing constructive journalism in AI-generated media can also encourage proactive public engagement, reinforcing findings that emphasize media’s role in enhancing community resilience \cite{kahneman1984choices}. These insights illustrate AI’s potential to create media that informs and enhances individuals' and communities' psychological and emotional resilience.

\subsubsection{Regulation is Needed to Prevent Unconscious Manipulation of Society and Individuals by AI-Generated Media.}

Our research shows that presentation style can significantly impact emotions and self-efficacy, even when facts remain unchanged, aligning with findings in journalism research \cite{wahl2020emotional}. This underscores risks like manipulation \cite{zuboff2019surveillance}, bias reinforcement \cite{nazar2020artificial}, and social tension exacerbation \cite{shah2022comprehensive} as AI-generated media becomes more prevalent.

Mitigating these issues requires robust governance frameworks. AI systems should be transparent, disclosing values, perspectives, and prompts behind content generation \cite{floridi2018ai4people}. Strict guidelines and oversight are essential to prevent misuse, especially in politics, corporate branding, and public opinion formation \cite{nishal2024envisioning}.

AI bias requires deeper examination. While we focused on emotional and efficacy impacts, data and algorithms can introduce biases. AI developers should use diverse datasets and implement real-time content monitoring. Interdisciplinary teams should regularly audit AI systems to ensure ethical standards, as suggested in prior works \cite{binns2018fairness,nishal2024envisioning}.

Moreover, AI-generated media can combat misinformation through fact-checking and balanced perspectives, as supported by Graves \cite{graves2018understanding}. Ethical use of AI-generated media demands transparency, regulation, and continuous oversight to maximize benefits while minimizing risks \cite{nishal2024domain}.

\subsubsection{Generative Podcasts Possess Broad Potential Future Applications Scenarios.}

AI-generated podcasts present significant opportunities in HCI, particularly for providing accessible media to visually impaired individuals for news, education, and navigation \cite{bigham2010vizwiz}. They can complement or even replace traditional text and graphical interfaces, addressing specific user needs.

AI-generated podcasts excel in multitasking environments. Users can receive tailored information without interrupting their activities. In professional settings, podcasts can serve as supplementary tools. The versatility makes podcasts a valuable secondary-task medium, enhancing graphical user interfaces (GUIs) with non-intrusive information access \cite{boulos2006wikis}.

In specialized fields like education and healthcare, AI-generated podcasts are valuable auxiliary tools. Healthcare professionals and educators can stay informed through customized content, while elderly or less literate users benefit from simplified information. By adjusting content complexity and style based on user needs, AI improves learning outcomes and delivery efficiency \cite{rawat2023exploring}.

\subsubsection{Towards AI-mediated Personalization of News Media}

AI’s distinctive value in podcast generation is its ability to adjust information dynamically. AI can aggregate and present news from multiple perspectives and flexibly modify the content's constructiveness. Unlike traditional producer-centered podcasts, AI-generated content can be more consumer-focused, providing personalized experiences based on individual needs, interests, and emotions \cite{wang2023democratizing}.

AI’s capacity to modulate information presentation can significantly impact social opinion and collective cognition. By refining how information is displayed, AI can shape public emotional responses and cognitive frameworks regarding events or issues. This personalization can significantly influence public opinion and societal cognition, aligning with studies on media framing and user engagement \cite{scheufele2007framing,sundar2010personalization}.

Future research should explore how AI can selectively gather and edit information while mitigating bias and avoiding information bubbles \cite{sharma2024generative}, ensuring a more comprehensive understanding of news events \cite{shrestha2024first}.

\subsection{Limitations and Future Work}

While this study provides insights into the potential of AI-generated podcasts, several limitations must be acknowledged. First, the relatively small sample size may have limited the statistical power and generalizability of our findings. A larger and more diverse sample would likely yield more robust results. Additionally, our study focused on a binary contrast between constructive and non-constructive content in AI-generated podcasts. This approach, while useful for initial exploration, oversimplifies the complexity of news and information presentation. Future research should explore a broader spectrum of content styles.

The study's controlled environment may not fully represent real-world conditions, where factors like multitasking, distractions, or biases could influence responses. The short-term focus leaves the long-term emotional and cognitive impacts unexplored.

Future studies should involve larger, more diverse samples and examine AI-generated content in naturalistic settings to improve ecological validity. Longitudinal research is needed to assess the long-term effects of AI-generated content on emotional health, cognitive processes, and behavior.

Finally, future research should explore the ethical dimensions of AI-generated content, including the potential for bias, misinformation, and manipulation. As AI technology evolves, establishing guidelines and safeguards is crucial to ensure responsible use that benefits society.

\section{CONCLUSION}

This work presents GenPod, a generative AI-driven pipeline that creates news podcasts in both constructive and non-constructive formats. By TTS technology, GenPod enables us to examine how different presentation styles influence listener perceptions. In a between-subjects study involving 65 participants, we used a mixed-methods approach, combining quantitative and qualitative analyses to assess the effects of these podcasts on listeners' emotions and self-efficacy. Our findings show that constructive podcasts significantly reduced negative emotions and enhanced self-efficacy compared to non-constructive podcasts. This research not only underscores the emotional and psychological impacts of AI-generated content but also provides valuable insights into the responsible and effective integration of such technologies into HCI designs while addressing potential ethical risks.


\bibliographystyle{ACM-Reference-Format}
\bibliography{sample-base}
\end{document}